\documentclass[useAMS,usenatbib]{mn2e}
\usepackage{epsfig}

\newcommand{\kms}{$\mbox{km~s}^{-1}$}
\newcommand{\lta}{\raisebox{-0.6ex}{$\,\stackrel
{\raisebox{-.2ex}{$\textstyle <$}}{\sim}\,$}}

\title[Annual Cycles in 1519--273 and 1622--253]{Annual Cycles in the Interstellar Scintillation Timescales of PKS\,B1519--273 and PKS\,B1622--253}
\author[Carter et al.]{Steven J.B. Carter$^{1}$\thanks{E-mail:Steven.Carter@utas.edu.au}, Simon P. Ellingsen$^{1}$, Jean-Pierre Macquart$^2$, James E.J. Lovell$^{1}$\\
$^{1}$School of Mathematics and Physics, University of Tasmania, Private Bag 37, Hobart, TAS 7001, Australia\\
$^{2}$Applied Physics, Curtin University of Technology, GPO Box U1987, Perth, WA 6845, Australia}
\begin{document}

\date{Accepted. Received ; in original form }

\pagerange{\pageref{firstpage}--\pageref{lastpage}} \pubyear{2009}

\maketitle

\label{firstpage}

\begin{abstract}
We have used the University of Tasmania's 30-m radio telescope at Ceduna in South Australia to
regularly monitor the flux density of a number of southern blazars.  We report the detection of an annual cycle in the variability timescale of the centimetre radio emission of PKS\,B1622--253.  Observations of PKS\,B1519--273 over a period of nearly two years confirm the presence of an annual cycle in the variability timescale in that source.  These observations prove that interstellar scintillation is the principal cause of inter-day variability at radio wavelengths in these sources.  The best-fit annual cycle model for both sources implies a high degree of anisotropy in the scattering screen and that it has a large velocity offset with respect to the Local Standard of Rest.  This is consistent with a greater screen distance for these ``slow'' IDV sources than for rapid scintillators such as PKS\,B0405-385 or J1819+3845.
\end{abstract}


\section{Introduction}

Active galactic nuclei (AGN) exhibit variability in their emission over all wavelength ranges and on a variety of different timescales.  At radio wavelengths the dominant emission from AGN is from relativistic jets of material ejected close to our line of sight.  Inverse Compton scattering limits the brightness temperature of the radio emission to be $\lta 10^{12}$K  \citep{KPt69}, which in turn sets a lower limit on the angular size of the radio emission region.  For most AGN the linear size of the radio emission region inferred from these arguments is of the order of light-weeks to light-months.  Hence causality arguments require that any variability in the radio emission from the source must have characteristic timescales of weeks-months or longer.   However, a significant number of AGN have been observed to vary at centimetre radio wavelengths on timescales of a few days or less \citep[e.g.][]{H84,Q+92,KC+01,L+03}.  If this variability cannot be intrinsic to the source, it must be a propagation effect imposed by the medium between the source and observer - interstellar scintillation (ISS).  \citet{L+08} have demonstrated that even when factors such as Doppler boosting are taken into account,  any AGN small enough to show intrinsic variability on a timescale of days must be of sufficiently small angular size to also exhibit interstellar scintillation.

The ionized interstellar medium (ISM) scatters the radiation from the source, and if the angular size of the source is comparable to the angular size of the density fluctuations in the ISM, interference effects produce variations in the intensity of the radio emission at the location of the observer.  Interstellar scintillation is typically modeled as occurring in a single, thin screen of turbulent plasma \citep{R90,N92}.  The relative transverse motion of the screen and the Earth means that that the modulated intensity pattern observed at a particular location changes with time, causing the observed variability in the centimetre radio emission.  There are two definitive tests for ISS as the cause of radio variability: an annual cycle in the variability timescale and a time delay in the intensity variations observed at two widely separated telescopes.

The variability timescale is set by the relative transverse motion of the Earth and the scattering screen.  As the Earth orbits the Sun the projection of its velocity vector onto the scattering screen will change throughout the year and the change in the relative transverse velocity is around 30~\kms .  Provided that the scattering screen does not have a large intrinsic transverse velocity this will result in significant changes in the variability timescale over the course of a year, which will repeat annually: an annual cycle.  Annual cycles have been measured in a handful of AGN: B0917+624 \citep{JM01,R+01}, J1819+3845 \citep{DTdB03}, PKS\,B1257-326 \citep{B+03}, and PKS\,B1519--273 \citep{J+03} J1128+5925 \citep{G+07}.  The spatial scale of the intensity pattern is typically much greater than the size of the Earth and so the intensity variations observed at widely separated telescopes are highly correlated.  However, if the separation of the telescopes is such that a change in intensity can be detected on timescales less than the time the changing intensity pattern takes to travel between the telescopes (which is a function of the projection of the relative velocity vector onto the baseline vector and the baseline length), then a time-delay in the intensity patterns can be observed.  Time-delays have also been measured in a handful of AGN: PKS\,B0405-385 \citep{J+00}, J1819+3845 \citep{DTdB02} and PKS\,B1257-326 \citep{B+06}.  Measurements of ISS induced variability in radio sources can be used to study both the source and the interstellar medium, through the technique of Earth-Orbit Synthesis \citep{MJ02}.  From the observations made to date, the distinguishing characteristic of the most rapidly varying sources appears to be that the scattering screen is located close (within a few 10s of pc) to the Earth.

Most of the heavily studied scintillating AGN have characteristic timescales for the intensity variations of the order of a few hours.  This means that in most cases the timescale can be measured to reasonable accuracy in observations of duration 12 hours or more.  The largest survey to date for ISS in AGN is the Micro-Arcsecond Scintillation-Induced Variability (MASIV) Survey \citep{L+03} where the VLA was used to monitor 443 compact flat spectrum radio sources in four epochs of 3 to 4 days each at 5~GHz. Variability was detected in 56\% of sources but only 12\% showed ISS at every epoch and the great majority of variability timescales were of order 1 day or longer \citep{L+08}.   The more typical scintillating AGN therefore shows variability on longer timescales and does so intermittently.   For a source whose angular size $\theta_{\rm src}$ exceeds the Fresnel angle (i.e. $\theta_{\rm src} > \sqrt{\lambda/2 \pi D}$, where $\lambda$ is the wavelength of the observations and $D$ the distance from the Earth to the scattering screen), the size of the scintillation pattern is governed by the source size, and has a physical scale $D \theta_{\rm src}$ at the Earth, so that the timescale on which the pattern advects across an observer is $T_{\rm char} = D \theta_{\rm src}/V_{\rm trans}$ (where $V_{\rm trans}$ is the relative transverse velocity of the screen and observer).  Hence the observed scintillation timescale increases linearly with the effective distance to the scattering material.  \citet{L+08} have also shown that the scintillation timescales observed for the majority of sources in the MASIV sample is inconsistent with the variability being produced in a screen moving with the local standard of rest.  So the observation of  long timescales combined with evidence that the majority of screens that cause ISS are not moving with the local standard of rest, together suggest that scattering is likely occurring at distances from Earth of many hundreds of parsecs in most cases.

The long characteristic timescales prevalent in the majority of IDV sources presents some problems for their study.  It makes measurement of a time delay difficult or impossible, since the time taken for an inflection in the intensity is large compared to the time taken for the interference pattern to travel the distance between the telescopes.  While, characterising the timescale of the variations, and how that changes throughout the year requires near continuous monitoring.  The high subscription rates for large interferometer arrays where most previous IDV studies have been conducted (VLA, ATCA, WRST), are such that it is not possible to get sufficient time on these facilities to undertake such a project.  The COSMIC project \citep{M+05} has been designed specifically to fill this niche, using a 30-m radio telescope to undertaken high-cadence monitoring of a small number of strong, longer timescale IDV sources.  Here we report the results of the first two years of monitoring for two of the COSMIC sources PKS\,B1519--273 and PKS\,B1622--253.

\section[]{Observations \& Data Reduction}

We have used the University of Tasmania 30-m radiotelescope near Ceduna, South Australia to monitor the intensity of PKS\,B1519--273 and PKS\,B1622--253 at 6.7~GHz for a period of approximately two years.  The observations were made as part of the COntinuous Single-dish Monitoring of Intraday variability at Ceduna (COSMIC) project.  The motivation, background and observation techniques for the project are described in detail in \citet{M+05}.

\begin{figure*}
  \psfig{file=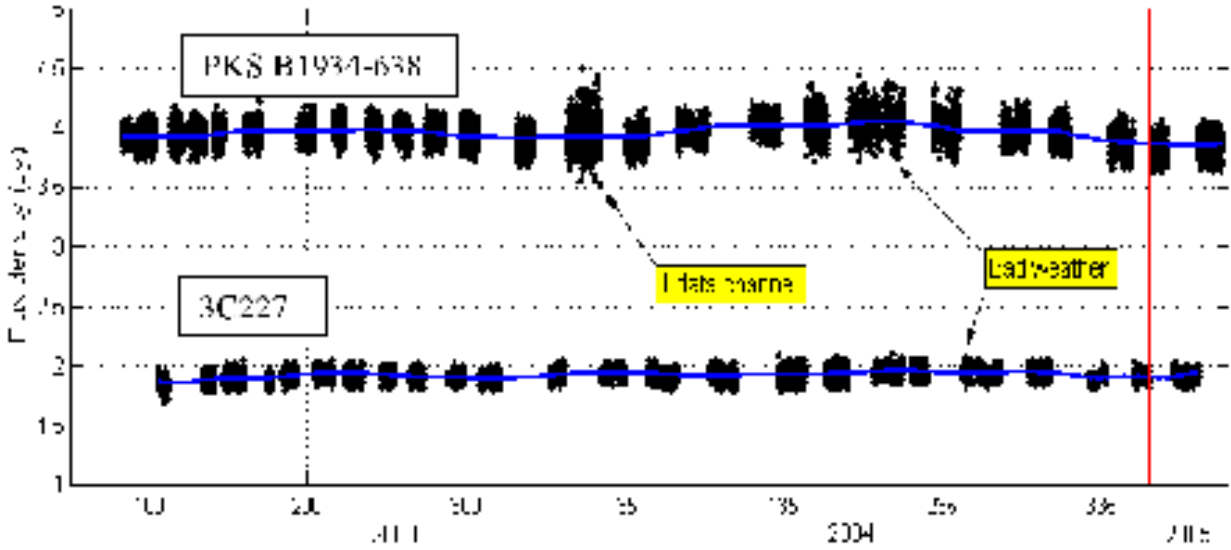,height=0.425\textwidth}
  \caption{Intensity of the COSMIC calibrators over the period 2003 - 2005.  The extremes in the scatter of individual flux density measurements can be seen from the spread in the measured intensity, as can the slow changes in the mean intensity, believed to be due to thermally induced variations in the output power of the calibration noise diode.  The red lines mark the start of the 2004 and 2005 calendar years, and a number of periods of poor data quality are labelled.}
  \label{fig:calibrators}
\end{figure*}

The intensity of PKS\,B1519--273 and PKS\,B1622--253 were measured using a series of four scans, two driving the telescope in right-ascension and two driving the telescope in declination.  Observations of these two sources were interleaved with observations of the calibrator 3C227 and another variable source AO\,B0235+164 \citep{S+08}. The measured intensities were corrected for antenna pointing and gain-elevation effects using the data reduction methods described in detail in \citet{M+05}.  The sources observed in the COSMIC project are split into two groups, those with declinations north of the zenith (approximately --32$^\circ$) at Ceduna and those which pass south of the zenith.  We observe one group of sources for periods of around 10-15 days and then swap to the other group.  Figure~\ref{fig:calibrators} shows the intensity of the two COSMIC calibrators (3C227 for the northern sources and PKS\,B1934-638 for the southern sources) over the period March 2003 - January 2005.  Prior to each scan a noise diode was switched and compared to a 1 dB step.  Comparison of the height of the noise diode to a source of known (or assumed) flux density then allows conversion of the intensity measurements from VFC (voltage to frequency converter) counts into Jy.  Figure~\ref{fig:calibrators} shows the results of applying a single conversion factor, determined from the first month of observations of PKS\,B1934-638 to the entire dataset.  Any changes in the amplitude of the noise diode will then cause apparent variations in the calibrator sources and this is the cause of the small long-timescale variations observed in Figure~\ref{fig:calibrators}.  The mean flux densities of calibrators PKS\,B1934-638 and 3C227 measured from this data are 3.952 $\pm$ 0.093~Jy and 1.925 $\pm$ 0.048~Jy respectively, agreeing with the known calibrator flux densities at 6.7 GHz of 3.92~Jy \citep{R94} and 1.90 Jy \citep{B+77} respectively.  Figure~\ref{fig:calibrators} shows a polynomial fit to the daily mean intensity of PKS\,B1934-638 and 3C227.  These fits are used to determine the correction factor necessary to scale the mean flux density (for each observing block) to the assumed flux density for each calibrator.  This correction factor never exceeds 3.5 percent, and the measured RMS values for PKS\,B1934-638 and 3C227 after its application are 0.080~Jy and 0.042~Jy respectively.

\begin{figure}
  \psfig{file=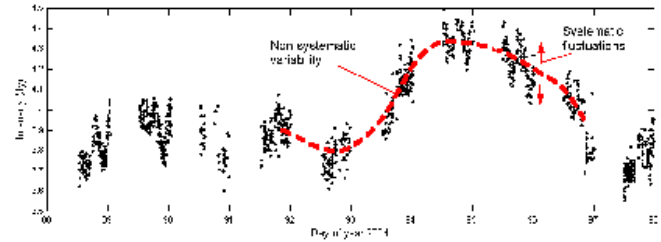,height=0.18\textwidth}
  \caption{The intensity variations observed in PKS\,B1622--253 over the period 2004 DOY 88 - 98.  Systematic variations of amplitude $\pm$ 0.15 Jy are present on timescales less than one day, however, the ISS induced variations are larger amplitude and longer timescale.}
  \label{fig:systematic}
\end{figure}

\begin{figure*}
  \psfig{file=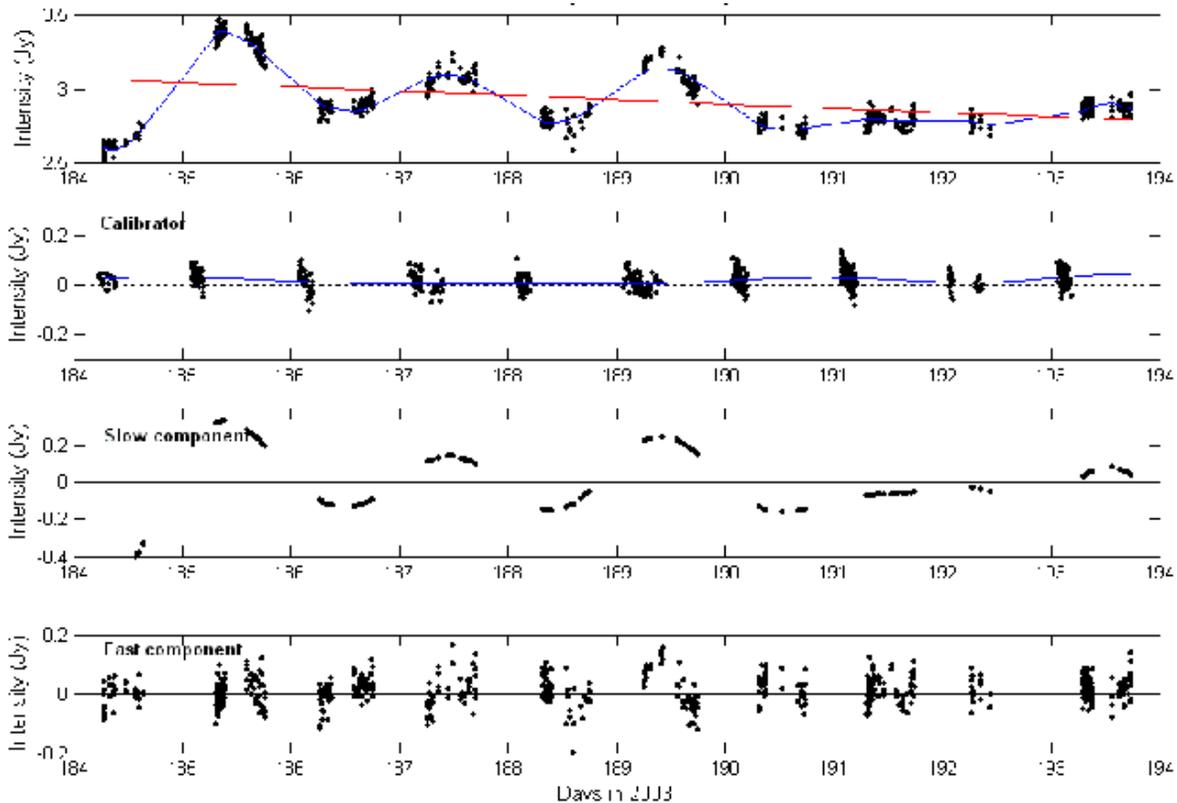,height=0.6\textwidth}
  \caption{An example of the polynomial filtering used to remove systematic variations from PKS\,B1519--273 and PKS\,B1622-235 data.  The top plot shows 10 days of raw data for PKS\,B1622--253, the blue line is the fitted low-pass polynomial filter (14th order polynomial in this case) and the red line a first order polynomial, which is fitted to the data to estimate intrinsic variations in the mean flux density of the source over the observing period.  The second plot shows the mean-subtracted calibrator (3C227) data for the same period, to which the same order polynomial has been fitted as for PKS\,B1622--253, this is used to estimate the systematic changes on timescales longer than a day.  The third plot shows the post-filtered, mean-subtracted flux density measurements for PKS\,B1622-253.  It is this data which is used for the time-series analysis.  The bottom plot shows the difference between the raw and filtered data, demonstrating the presence of systematic variations of amplitude $\sim \pm$0.15 Jy on timescales shorter than a day.}
  \label{fig:poly_filter}
\end{figure*}

The changes in the amplitude of the noise diode appear to be correlated with the ambient temperature.  Such changes occur on a wide range of timescales, not just the longer timescales apparent in Figure~\ref{fig:calibrators} and we do observe smaller amplitude ($\sim \pm$0.15 Jy) variations in the calibrators on diurnal timescales (Figure~\ref{fig:systematic}).  A number of approaches have been tested to try and correct for this effect, however, because the noise diode is housed within the telescope structure there is a variable time lag between a change in the ambient air temperature (which is measured), and the subsequent change in the temperature of the air around the noise diode.  The temperature of the noise diode is now measured, but was not at the time of the observations reported here.  During the summer months air conditioning is used to reduce excessive ambient air temperature variations around the receiver and back-end systems, but the need for proper temperature stabilization of these areas is now recognised.  The systematic variations in the calibrator intensity on timescales longer than a day are determined by normalizing the flux density, by dividing by its mean over the observing period and then fitting a polynomial to the data.  The order of the polynomial is approximately the same as the number of days in the observing period, so this approach is only able to correct for systematic variations on timescales of the order of a day or longer.  To remove the shorter timescale (higher frequency) systematic variations a polynomial low-pass filtering process is applied to the target-source data in the time domain.  An example of this process is shown in Figure~\ref{fig:poly_filter} for 10 days of PKS\,B1622-253 data.  An in depth description of the method, including the steps which are undertaken to make sure that the polynomial fit behaves at the two ends of the time range, is given in \citet{C08}.  This two-pass approach to removing/reducing the systematic variations in the data has been extensively tested using both real and simulated data and demonstrated to be both robust and effective.  Approaches which use the measured ambient temperature and a model of the changes in the noise diode power with temperature have also been investigated and, while they yield similar results, they are significantly more difficult to implement.  Ideally the calibrator measurements would be used directly to measure the variations in the target sources, however, the different right ascension of the calibrator 3C227 (9 hours right ascension) and PKS\,B1519--273, PKS\,B1622--253 means there is very little overlap and this approach cannot be used.

We have compared the observed scatter in the calibrator sources (after correction for systematic effects), with expectations from the radiometer equation and find that they are significantly higher.  Such effects are common in continuum measurements made with single-dish radio telescopes and are thought to be due to small and rapid variations in the electronic gain of the system.  There is no easy way to fix this problem, indeed it is one of the fundamental reasons that interferometers are superior for radio intensity measurements - the electronic gain variations are independent for a pair of antennas, hence they do not correlate and interferometers are typically able to achieve close to their theoretical sensitivity limits.  We observe deviations from thermal noise behaviour on timescales longer than 0.1 seconds and the measured uncertainty in our flux density measurements on a timescale of 5 seconds (approximately the time it takes to scan across the source) is approximately 2.5 times the theoretical radiometer estimates.  As outlined by \citet{M+05} the accuracy of the flux density measurements is a function of the intensity, but for PKS\,B1519--273 and PKS\,B1622--253 over the period 2003 - 2005 it was typically between 1 and 2 percent of the mean flux density ($\sim$ 30~mJy). 
  
\begin{figure}
  \psfig{file=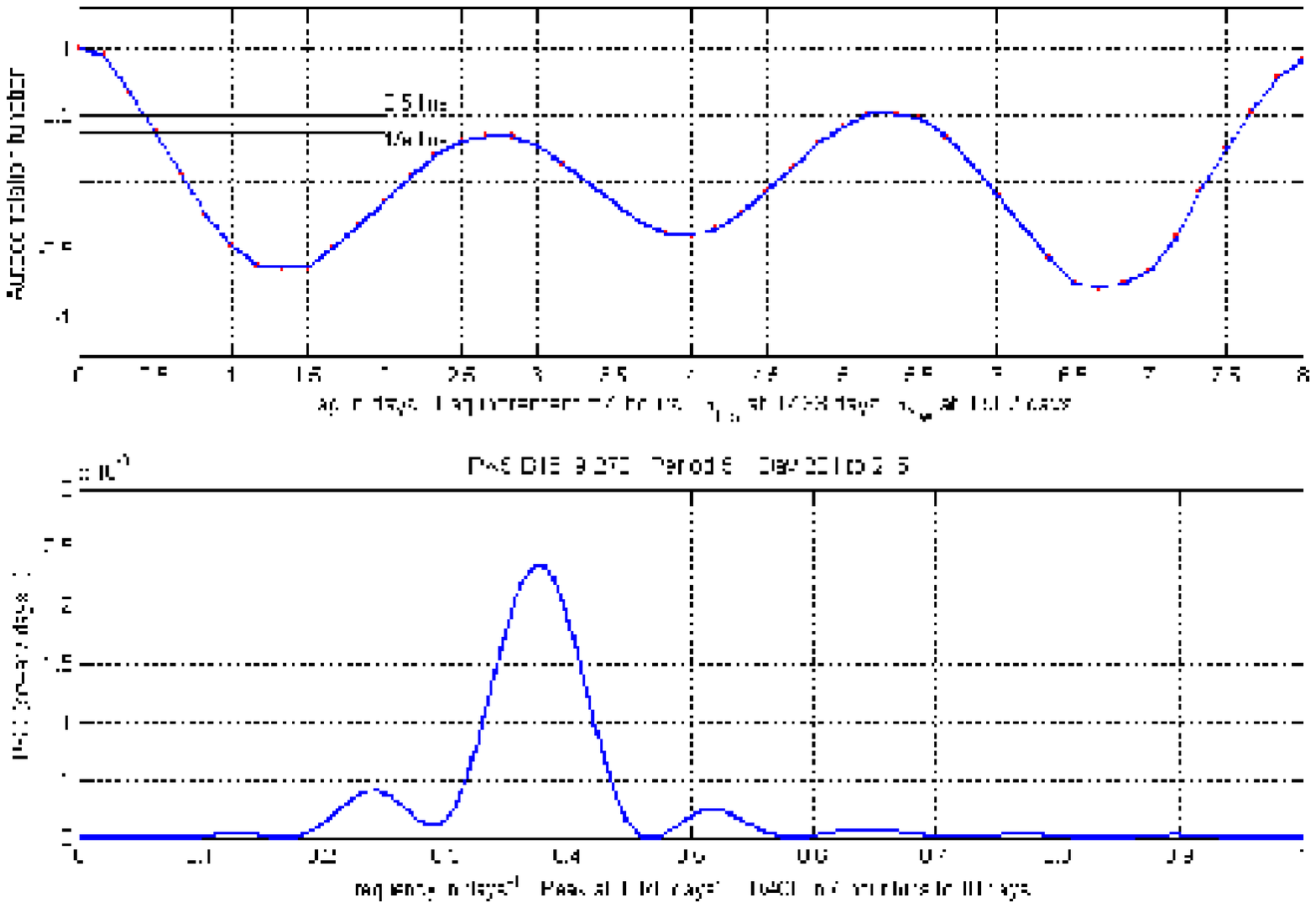,height=0.325\textwidth}
  \caption{The discrete auto-correlation function (DACF) and power spectral density (PSD) for PKS\,B1519--273 for the session 2003 DOY 204 - 215.  The PSD shows a clear peak at a timescale of 0.38 days$^{-1}$, corresponding to a quasi-period in the variations of approximately 2.6 days.}
  \label{fig:PSD}
\end{figure}

The longer timescale of the variations in PKS\,B1519--273 and PKS\,B1622--253 and the use of a single-dish radio telescope rather than an interferometer has required the utilisation of different methods of measuring the variability timescale than those most commonly used for IDV studies.  Most determinations of the characteristic timescales for IDV sources calculate the discrete auto-correlation function \citep{EK88} or DACF, and then measure the width at which the correlation drops to a half ($\tau_{0.5}$),or $1/e$ ($\tau_{1/e}$).  This approach does not work very well for the COSMIC data  due to the regular sampling (the source rise/set time at the telescope imposes a regular gap in observations each sidereal day).  When the characteristic timescale of the variations is close to a multiple of a day the artifacts in the DACF make it difficult to accurately estimate.  The Fourier Transform of the DACF is known as the power spectral density (or PSD) and this is much less affected by the artifacts than the DACF.  In most cases the PSD shows a single dominant peak (see for example Figure~\ref{fig:PSD}) and the corresponding timescale is the period we have taken as characteristic of the variability in this study.

Interstellar scintillation is an inherently stochastic process and so for some observing periods there is no well-defined quasi-period for the timescale of the variations.  As a cross-check on the characteristic timescale obtained from the PSD we used two secondary approaches to estimate the variability timescale.  The first cross-check was to systematically fold the data at a range of different periods and then fit a simple sinusoid to the folded data.  The period for which the amplitude of the fitted sinusoid is a maximum and the residuals to the fit are a minimum is then taken as an estimate of the quasi-period.  The second cross-check (which we refer to as scintle-counting), was to locate by eye the times of the turning points in the intensity variation time series and then estimate the quasi-period by dividing the duration between the first and last turning points by the number of turning points.  Only observing sessions for which the PSD, data-folding and scintle-counting all returned consistent estimates for the quasi-period were included when determining if the characteristic timescale of the variations exhibits an annual cycle in these sources.

\section[]{Results}

\begin{figure*}
  \psfig{file=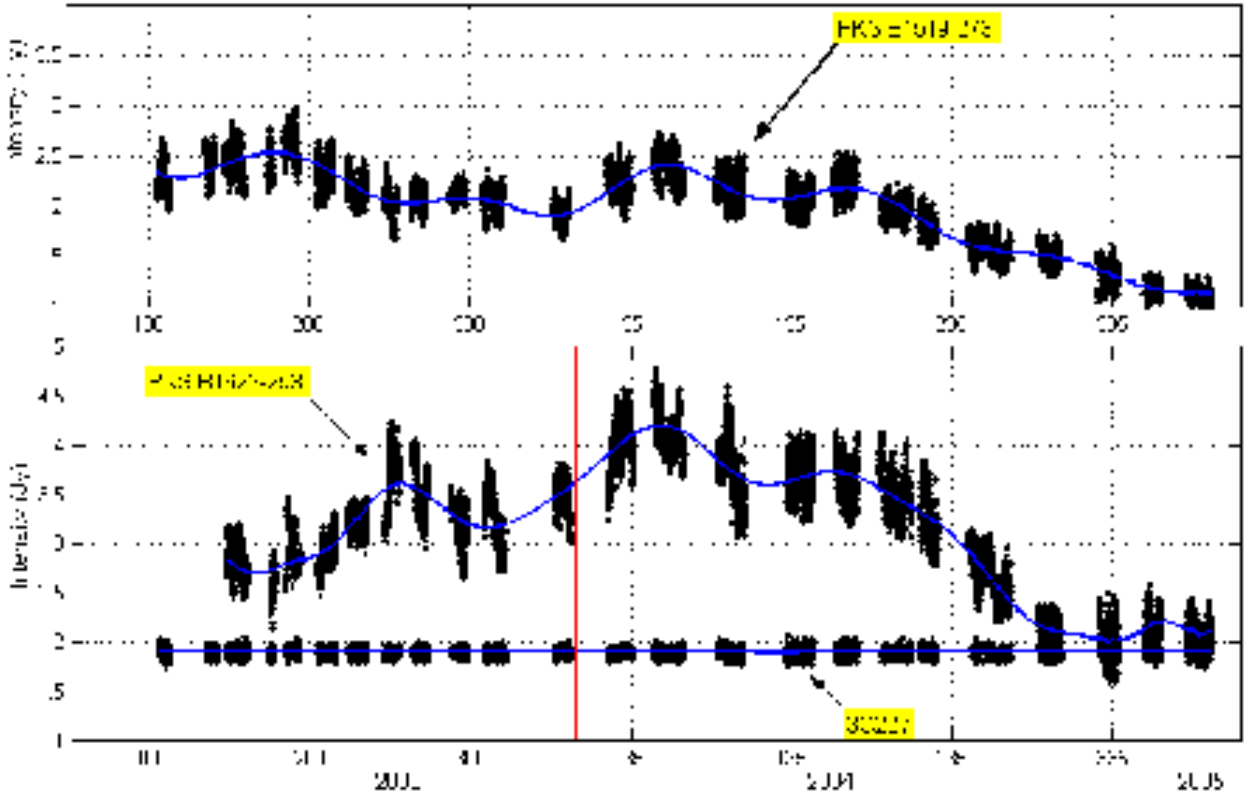,height=0.6\textwidth}
  \caption{Julian-day Intensity Monitoring (JIM) plot of PKS\,B1519--273 and PKS\,B1622--253 over the period 2003 - 2005.  Large changes in the mean flux density, due to intrinsic changes in the intensity of the sources are apparent on timescales of months and years in both sources, while the calibrator 3C227 shows no such variations.  The larger scatter apparent in the data for PKS\,B1519--273 and PKS\,B1622--253 compared to the calibrator is primarily due to short timescale ISS-induced variability which is compressed by the time-axis on this plot.}
  \label{fig:JIMplot}
\end{figure*}

\begin{figure}
  \psfig{file=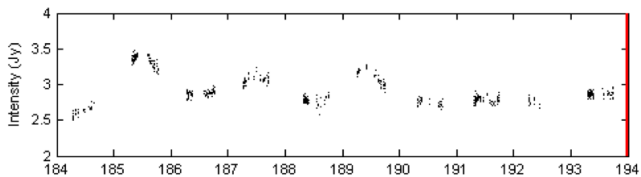,height=0.135\textwidth}
  \caption{The intensity of PKS\,B1622--253 versus time for a typical ten-day period.  In this case 2003 DOY 184 - 194.}
  \label{fig:1622_zoom}
\end{figure}

Figure~\ref{fig:JIMplot} shows the Julian-day Intensity Monitoring plots (JIM-plots) for PKS\,B1519--273 and PKS\,B1622--253 over the period 2003$-$2005.  The mean flux density of PKS\,B1519--273 at 6.7~GHz is seen to vary between about 2.5~Jy in mid-2003 and again in mid-2004, down to approximately 1~Jy by early 2005.  PKS\,B1622--253 shows even larger variations in mean flux density, peaking in excess of 4~Jy in early 2004 down to around 2 Jy later the same year.  The variability observed on timescales of months to years is believed to be intrinsic to the source due to evolution in the parsec scale jets.  Variations of this type are commonly observed in the centimetre wavelength emission of AGN.  It is also noticeable that the scatter in the measured intensity for PKS\,B1519--273 and PKS\,B1622--253 is significantly greater than observed for 3C227.  This is due to the ISS-induced short timescale variations, which are compressed by the large time range displayed in Figure~\ref{fig:JIMplot}.  Figure~\ref{fig:1622_zoom} shows a typical ten-day period for PKS\,B1622--253, where the character of the shorter timescale variability can be clearly seen.

\begin{figure}
  \psfig{file=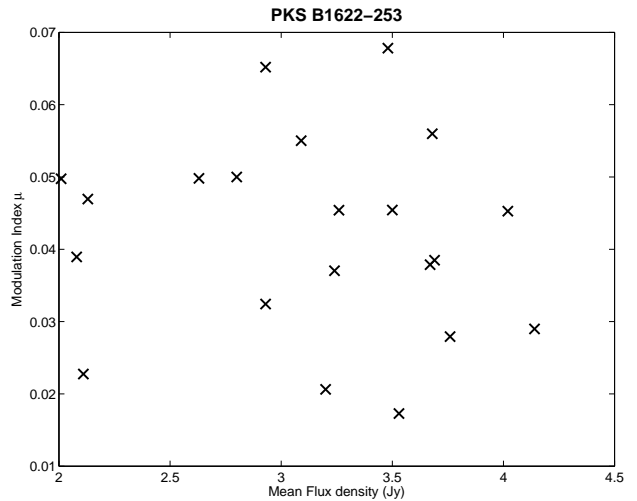,height=0.4\textwidth}
  \caption{There is no correlation between the modulation index and mean flux density in the PKS\,B1622-253 data.}
  \label{fig:1622_deltaSvS}
\end{figure}

The modulation indices ($\mu = \frac{\sigma}{S}$) for PKS\,B1519--273 and PKS\,1622--253 over the two year observing period were in the range 0.02 - 0.07 on timescales of 10 days, after correction for systematic effects ($\sigma$ is the standard deviation in the measured flux density and $S$ is the mean flux density over the observing period in question). This represents a lower-limit to the modulation index for the scintillation as some of the emission is expected to be from source components which are too large to exhibit ISS.  We investigated how the modulation index changed as the mean flux density of the sources changed and found a loose correlation between the two quantities for PKS\,B1519--273, which suggests that a significant fraction of the change in the mean flux density was due to evolution of the compact, scintillating core.  In contrast for PKS\,B1622--253 there is no good correlation (Figure~\ref{fig:1622_deltaSvS}), suggesting that the flux density changes include significant contributions from non-scintillating components of the source.  

\subsection{Annual cycles}

\begin{figure*}
  \psfig{file=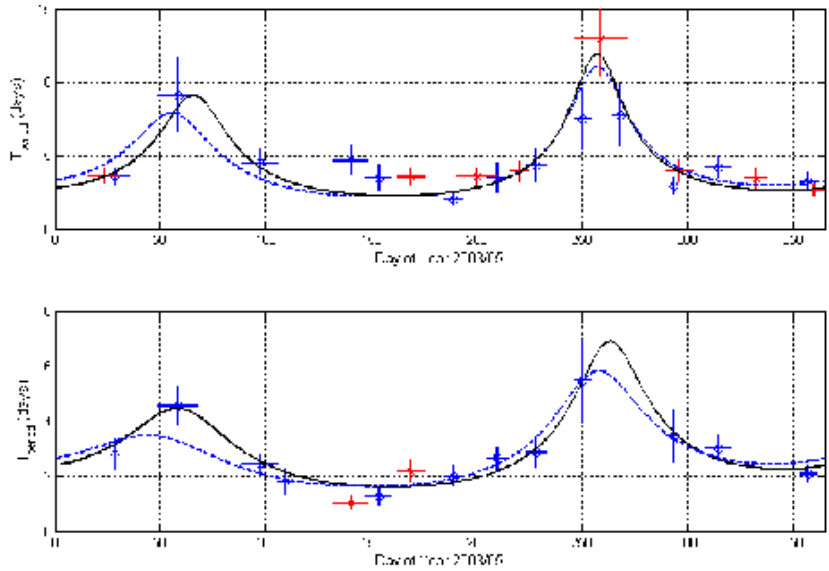,height=0.425\textwidth}
  \caption{Changes in the characteristic timescale as a function of day of year for PKS\,B1622--253 (top) and PKS\,B1519--273 (bottom).  The blue circle data points are from the first year of monitoring and the red cross data points are from second year.  All-parameter best fits shown as solid black lines.  Best fits for R=5 shown as dash blue lines.}
  \label{fig:annual_cycle}
\end{figure*}

Figure~\ref{fig:annual_cycle} shows the changes in the measured characteristic timescale as a function of the day of the year for both PKS\,B1519--273 and PKS\,B1622--253.  The blue circles are timescales measured during the first year of observations (2003), while the red crosses are from 2004.  Figure~\ref{fig:annual_cycle} clearly shows that the timescale of the variability in both of these sources changes with time in a manner which repeats on an annual cycle.  Such behaviour is an unambiguous indication that the short timescale variability observed in these sources is caused by interstellar scintillation.  The annual cycles displayed by the two sources are similar in their overall character as they are relatively close together on the sky and the ecliptic coordinates of the source play a major role in determining the times during the year when the variability slows and speeds up.  Although the two annual cycles are broadly similar, the actual timescales measured in each epoch typically differ by a factor of around two, with the variations in PKS\,B1519--273 generally being faster.  The observations used to measure the timescale in these two sources were interleaved and so the detection of different values demonstrates that they are not due to residual systematic effects.  Timescale analysis undertaken on the calibrators showed weak peaks in the PSD around periods of one day corresponding to diurnal thermal variations (and also weaker harmonics), but these contain much less power than the timescales measured for PKS\,B1519--273 and PKS\,B1622--253 during the same epoch.

\subsubsection{Estimating the error in timescale determination}

Prior to interpreting the apparent changes observed in the timescale of variations in PKS\,B1519--273 and PKS\,B1622--253 it is important to establish the uncertainty in our measurements.  The horizontal error bars in Figure~\ref{fig:annual_cycle} are just the length of the observing period.  The vertical (timescale) error bars were determined empirically from the distribution of scintle periods, normalised by the mean period for each observing period in order to remove the annual cycle effect.  This approach assumes quasi-periodicity, as discussed above.  It also assumes that the statistics of the normalised scintle distribution are constant over the whole data set (we will show in Section~\ref{sec:discussion} that this is a reasonable assumption).

Consider a given observing session (typically 10-15 days) which observes $N$ adjacent scintles with periods $T_1$ to $T_N$, and a mean scintle period, $T_{period}$, calculated from these $N$ values.  A period is measured as the scintle peak-to-peak or trough-to-trough time.  The $T_{period}$ estimate is associated with the middle of the observing period, with an uncertainty in time equal to the observing period.  The deviation of scintle period $T_i$ from $T_{period}$ is \[ T_{dev}^i = \frac{T_i - T_{period}}{T_{period}} \]

\begin{figure}
  \psfig{file=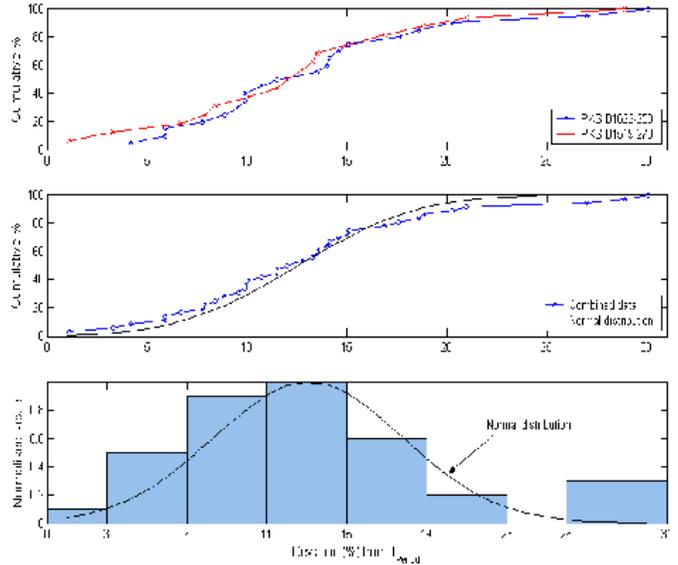,height=0.425\textwidth}
  \caption{The top plot shows the cumulative frequency distributions of $T_{dev}$ for each source.  The middle plot shows the cumulative frequency distribution of the combined $T_{dev}$ values, and the bottom plot shows the equivalent frequency distribution.  The smooth lines in the middle and bottom plots show Gaussian distributions of data with the same mean and standard deviation.}
  \label{fig:errorbar}
\end{figure}

This calculation accounts for the fact that $T_{period}$ may change between observing sessions either because of source evolution in structure and/or size (which leads to changes in the characteristic timescale $T_{\rm char}$), or because of the annual cycle (or a combination of the two).  Figure~\ref{fig:errorbar} shows distributions of $T_{dev}$ values for adjacent scintles in PKS\,B1622--253 and PKS\,B1519--273.  The distribution observed for both sources is very similar and the combined distribution is well represented by a normal distribution with a mean of 13.1 percent and a standard deviation of 6.7 percent.   In other words, half of all $T_{dev}$values calculated from two adjacent scintles are less than 13.1 percent, so if an observing period has a $T_{\rm char}$ value calculated from two scintles, an error bar of $0.13 \times T_{\rm char}$ corresponds to a confidence interval of approximately 50 percent that the true $T_{\rm char}$ value lies within the error bars.  Assigning a $\pm$ 20 percent error bar to the $T_{\rm char}$ values calculated from two adjacent scintles corresponds to a 2-$\sigma$ (95\%) confidence that the true $T_{\rm char}$ value lies within the error bars.  The corresponding error bars for $T_{\rm char}$ values calculated from 1, 3 and 4 scintles are given by Gaussian error theory as 28, 16 and 14 percent respectively.

The error bars for the timescale axis for each point were calculated using the method outlined above, except for points where the observed $T_{dev}$ for an observing period exceeded the expected error estimate, in which case the measured $T_{dev}$ was used.  

\subsubsection{Fitting an annual cycle model to the data}

The annual cycle in the timescale observed in ISS-induced variability can be used to study the scattering screen that produces it.  The relative transverse motion of the Earth about its orbit as viewed from the direction of the source is fixed for any particular source, however, there are five free parameters when fitting the observed annual cycle \citep{JM01} :
\begin{description}
  \item[$V_\alpha,V_\delta$:] The {\em screen velocity} measured in the right ascension and declination coordinate system (in \kms) compared to the Local Standard of Rest (LSR).
  \item[$s_0$:] The {\em characteristic scintle spatial scale}, is the coherence scale (in km) for the turbulence in the screen.  The characteristic scintillation timescale seen by the observer $T_{\rm char}$ is then \[ T_{\rm char} = \frac{s_0}{V_{trans}} \] where $V_{trans}$ is the relative transverse velocity of the screen and observer.
  \item[$R, \beta$:] Represents the {\em anisotropy} in the spatial pattern of the scintles.  The scintles are assumed to be elliptical in shape with an axial ratio $R$ and the major axis of the ellipse inclined at an angle $\beta$ with respect to the screen velocity vector.
\end{description}

We have undertaken a non-linear least squares fit of the annual cycle model to the data for PKS\,B1519--273 and PKS\,B1622--253.  The solid lines in Figure~\ref{fig:annual_cycle} are the optimal fits when all parameters are allowed to vary without restriction.  The values of these fits are summarised in Table~\ref{tab:annual_cycle}.  The line of best-fit passes within the error bars of most of the measured data points (the error bars showing the 95 percent confidence interval) and the reduced $\chi^2$ statistic for the two fits are 0.8 and 2.1 for PKS\,B1519--273 and PKS\,B1622--253 respectively.  Obtaining a $\chi^2$ value close to 1 for both source demonstrates that the error-bar estimation is reasonable.

The error estimates in Table~\ref{tab:annual_cycle} were obtained by investigating the variation of $\chi^2$ while 2 parameters were simultaneously varied.  This approach was taken as the effects of the different parameters on the annual cycle fit are not independent, indeed we found the fitting to be quite insensitive to the interplay between the anisotropy and LSR velocity parameters.  The 1-$\sigma$ error estimates given in Table~\ref{tab:annual_cycle} were obtained from the width of the $\chi^2 = 2.3$ error ellipse along the relevant axis, as is appropriate for systems with two degrees of freedom.  The interdependence of the different parameters in the model means that the annual cycle fitting is not a particularly effective way of estimating the anisotropy of the turbulent screen.  This is also demonstrated in Figure~\ref{fig:annual_cycle} where the black line shows the best fit when all parameters are allowed to vary and the blue-dashed line shows the fit when the anisotropy is restricted to be $R \leq 5$ (the best fit for which occurs when $R = 5$).  For both sources the best-fit models have an anisotropy ratio more than three times greater than this, however, the annual cycles in the two cases are not dramatically different.  The anisotropy can be more directly and accurately estimated from time-delay observations \citep[e.g.][]{B+06}, however, as outlined above these are not practical for the more slowly varying IDV sources we are dealing with here.  As the scintillation properties vary with wavelength in a fashion which is both significantly different from the differences expected due to changes in the source, and predictable, it is clear that simultaneous observations covering multiple frequencies would be greatly beneficial.  Observations of other sources have also seen more extreme variability in the polarized component of the flux density, which is interpreted as indicating the polarized emission arises from a smaller region than much of the total intensity emission \cite[e.g.][]{RKJ02}.   So observations at multiple frequencies and preferably also polarization information can greatly aid in disentangle source and screen properties and break some of the degeneracies present in annual cycle fits at a single frequency.

\begin{table*} 
\caption{Annual cycle model best-fit values.}
\begin{tabular}{llllll} \hline
{\bf Source} & $R$       & $\beta$ (rad) & $V_\alpha$ (\kms) & $V_\delta$ (\kms) & $s_0$ ($10^6$ km) \\ [2mm] \hline
PKS\,B1622--253 & $18.0 \pm 4.0$ & $0.17 \pm 0.02$ & $-37.5 \pm 3.0$ & $8.4 \pm 0.5$ & $9.2 \pm 0.9$ \\
PKS\,B1519--273 & $15.1 \pm 3.0$ & $0.07 \pm 0.02$ & $-52.0 \pm 7.0$ & $12.1 \pm 1.0$ & $7.1 \pm 0.4$ \\
\end{tabular} \label{tab:annual_cycle}
\end{table*}

\section[]{Discussion} \label{sec:discussion}

\subsection{PKS\,B1519--273}

The BL Lac object PKS\,B1519--273 is listed in the first Parkes catalogue, with flux densities of 2.0 Jy at 2.7 GHz and 2.3 Jy at 5 GHz \citep*{BSW75}.  It is a point source to most instruments at most wavelengths, although both the VLBA at 2-cm \citep{LH05} and the VSOP 5~GHz survey \citep{D+08} show it to be slightly extended.  There are no confusing radio sources nearby, and it is a weak source of soft X-rays \citep{U+96} and possibly of $\gamma$-rays \citep{F+94}.  Optically, PKS\, B1519--273 is identified as a B-filter magnitude 17.7 star-like object with a featureless spectrum \citep{VCV03}. Visual observations have not identified a host galaxy, because it is obscured by nearby and partly overlapping objects.  PKS B1519--273 was included in high resolution I-band studies of 24 BL Lac objects carried out using several telescopes between late 1999 and early 2001, and found to lie at a redshift of z = 1.294 based on 4,000 - 10,000 \AA\ low-resolution spectral observations \citep{H+04}.  A similar redshift was obtained by \citet{S+05}. In both cases these measurements are based on a single emission line in the spectrum identified as Mg{\sc ii} and are very different from the value of 0.07 given by \citet{Z+02}, although it is not clear how or where that measurement was obtained.

PKS\,B1519--273 is well known to display pronounced intrinsic radio variability on timescales of months. The ATCA blazar monitoring program of \citet{B03} included PKS\,B1519--273, and eleven sets of flux density observations at 8.6 GHz showed it to vary between a mean flux density of 1.8 and 2.2 Jy over a 12 month period (starting in February 2001).  Radio variability on time scales of days was discovered in PKS\,B1519--273 at 4.8 and 8.6 GHz during an ATCA survey in 1994 \citep{KC+01}, and was also observed in the 1997-2001 ATCA blazar monitoring program \citep{B03}.  Multi-frequency ATCA observations of PKS\,B1519--273 were made for five consecutive days starting on 10 June 1996, and again starting on 9 September 1998 \citep{M+00}.

Multi-frequency ATCA observations of PKS\,B1519--273 have found that the transition frequency between strong and weak scattering is in the vicinity of 4.8 GHz \citep{M+00}, so the COSMIC observations at 6.7 GHz should be in the weak scattering regime.  Weak scattering is a broadband effect, and \citet{M+00} found that the light curves observed at 8.6 and 4.8 GHz were well correlated.  There was also some correlation in their observations of light curves between 4.8 and 2.5 GHz, but there was clearly a significant change in the nature of the concurrent scintillation between 2.5 and 1.4 GHz.

The annual cycle in PKS\,B1519--273 confirms the work of \citet{J+03}, who concluded that this source displayed an annual cycle in its variability time scale based on an examination of ATCA data over ten epochs, including the five days in 1998 data; and seven epochs in 2001, each with 2-3 days of data.  Most of these observing periods were too short to record more than a single scintle, and the characteristic variability of each epoch was thus measured as $\tau_{0.5}$, calculated from the DACF.  Three of the ten data points span days 120 to 150 of the year, with $\tau_{0.5}$ values of $\sim$3 hours (8.6 GHz) and $\sim$5 hours (4.8 GHz).  For quasi-periodic scintles, $T_{\rm char} \approx 6 \times \tau_{0.5}$ so at the Ceduna observing frequency of 6.7 GHz the expected value of $T_{\rm char}$ is approximately 6 $\times$ 4 hours = 24 hours during this time of year.  This agrees well with the values observed by Ceduna, as can be see in Figure~\ref{fig:annual_cycle}.

\subsection{PKS\,B1622--253}

PKS\,B1622--253 is listed in the first Parkes catalogue, with flux densities of 2.3 Jy and 2.0 Jy at 2.7 GHz and 5 GHz respectively \citep{BSW75}.  It is optically very faint and lies behind the Ophiuchus Cloud, a dense interstellar gas cloud south of $\rho$ Ophiuchus \citep{H+94}.  PKS\,B1622--253 has a redshift of z = 0.786 \citep{dSA+94} and an unresolved optical counterpart of B-filter magnitude 21.9 (Schlegel et al., 1998).  \citet{P+05} present VLA radio images of PKS\,B1622--253 made in June 2002, at 3.6 and 6~cm. The images show extended structure consistent with the core-jet morphology of the central engine model.  The main features are a powerful radio core, FR II radio lobes consisting of a diffuse jet-like eastern extension, and a western lobe that is also part of a jet. $\gamma$-ray emission from PKS\,B1622--253 was detected by the Energetic Gamma-Ray Experiment Telescope (EGRET) on board the Compton gamma ray observatory \citep{N+96}.  From the EGRET observations, \citet{H+94} tentatively determined the $\gamma$-ray spectrum of PKS\,B1622--253 to be described by a power law with spectral index -1.9 $\pm$ 0.3.  However the $\gamma$-ray flux is PKS\,B1622--253 is variable, and changed by a factor of $\sim$2 over the 1991-93 EGRET observation period.

\citet*{T+98} noted that while the Parkes 90 catalogue records PKS B1622--253 as having a flux density of 2.02 Jy at 5 GHz, the subsequent Parkes-MIT-NRAO survey records its flux density as 3.5 Jy at 4.85 GHz.  Their ATCA observations observed its flux density at 8.6~GHz to decrease from 2.5 Jy to 0.9 Jy in the 10 month period prior to July 1996.  Similarly, 22 GHz observations with the VLBA \citep{J+01} found the flux density of the core to vary by a factor of five over the course of a year.  Radio variability on time scales of days was discovered in PKS B1622--253 by the ATCA blazar monitoring program from July 1997 to January 2002 \citep{B03}.  The source was excluded from the earlier 1994 ATCA survey because it failed to meet a compactness selection criterion.  The source also displays circular radio polarisation variability \citep{T+03}, but this is too weak to be observed by the Ceduna telescope.

PKS\,B1622--253 is also assumed to be in the weak scattering regime at 6.7 GHz.  This has not yet been conclusively determined through multi-frequency observations, but the scintillation characteristics of this source are similar to PKS\,B1519--273, and the flux density variability time scales of the two blazars follow similar annual cycles.  The annual cycle in PKS\,B1622--253 has not been previously reported in the literature.  The annual cycle we observe is clearly very similar to the annual cycle in PKS\,B1519--273, although the characteristic timescale in this source is typically significantly longer.

\subsection{Anisotropy inferred from annual cycle measurements}

For both sources the optimum fit to an annual cycle implies highly anisotropic scintles and large LSR velocity offsets, particularly $V_{\alpha}$.  This is unsurprising as significant velocity offsets are expected for scintillation with relatively long variability time scales (days rather than hours), which are likely to be associated with more distant scattering screens that typically lie hundreds of parsecs from Earth, and thus are in motion with respect to the LSR.  Anisotropic scintles have been associated with other blazar observations, notably PKS\,B0405-385 \citep*{RKJ02} and J1819+3845 \citep{MdB06}.  Many authors have argued that, for scintillation due to scattering by the ISM, electron density fluctuations are expected to be elongated along the local direction of the magnetic field \citep[e.g.][]{GS95,CB02}.  Alternatively, anisotropy may be due to an anisotropic source structure, however, further observations at multiple radio frequencies are required to disentangle the source and screen properties.

\begin{figure*}
  \psfig{file=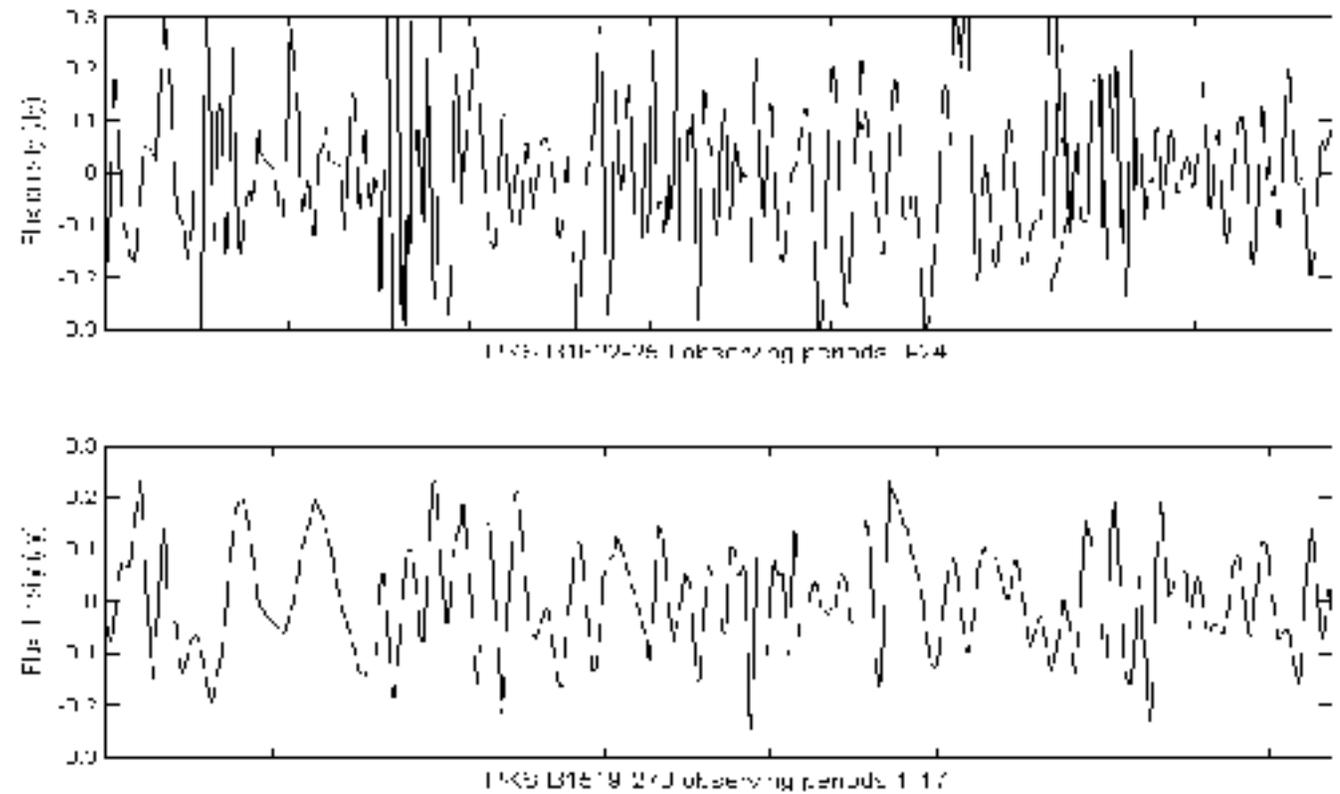,height=0.6\textwidth}
  \caption{PKS\,B1622--253 (top) and PKS\,B1519--273 (bottom) data adjusted by the time scales calculated from the best fit annual cycle models to remove the annual cycle effect, with the modified time series adjusted to ``constant'' variability timescales of 4.5 days and 3.0 days respectively.}
  \label{fig:annual_removed}
\end{figure*}

The annual cycle effect can be removed from a flux density time series by using the best-fit models for annual cycles in $T_{\rm char}$ to correct the data for departures from a specified $T_{\rm char}$ value.  This enables examination of the overall pattern of scintillation modulations throughout the 2003 -- 05 observing campaign, without the effects of ISM scattering material motion and anisotropy.  This parallels the exercise carried out by \citet{DTdB03} for J1819+3845 data spanning two years of observations.  Figure~\ref{fig:annual_removed} shows all the PKS\,B1622--253 and PKS\,B1519--273 data for 2003 -- 05 with the annual cycle effect removed by adjusting the data variability to $T_{\rm char}$ values of 4.5 and 3.0 days respectively.  The reference $T_{\rm char}$ values are close to the mean $T_{\rm char}$ values for the annual cycle, as can be seen from Figure~\ref{fig:annual_cycle}.  Figure~\ref{fig:annual_removed} is very similar to Figure 12 in \citet{DTdB03}.  As was found for J1819+3845, the variability of both PKS\,B1622--253 and PKS\,B1519--273 is, by eye, similar throughout the observing campaign. In other words, just as for J1819+3845, the scintle statistics did not change significantly over the two years of  observations for PKS\,B1622--253 and PKS\,B1519--273.

\section{Conclusions}

We have detected an annual cycle in the characteristic timescale of the variability in the radio emission of PKS\,B1622--253 and confirm the annual cycle in PKS\,B1519--273 determined on the basis of a much smaller and more sparsely sampled dataset.  The characteristic timescale of the variability in these sources ranges from around one day (PKS\,B1519--273 at the fastest time of the year) to around 10 days (PKS\,B1622--253 at the slowest time of the year).  These findings demonstrate the unique range of parameter space being investigated by the COSMIC project.  Such long characteristic timescales cannot be practically monitored using premier interferometer systems such as the VLA or ATCA because these instruments are in too high demand to dedicate long periods of time to flux density monitoring.  The best-fit annual cycle for both these sources implies scattering screens with significant transverse velocities with respect to the LSR and a high degree of anisotropy.  Through the COSMIC program, we will continue to monitor these and other strong southern IDV sources to look for signs of changes in the source structure and scattering screen.

\section*{Acknowledgments}

We would like to thank P.M. McCulloch and D.L. Jauncey for their efforts in making the Ceduna radio telescope the radio astronomy facility it is today, and in the planning of the COSMIC project.  We would like to thank Bev Bedson for her tireless efforts in keeping the Ceduna telescope operating.  Financial support for this work was provided by the Australian Research Council.  This research has made use of NASA's Astrophysics Data System Abstract Service.

\label{lastpage}

\end{document}